\documentstyle[aps,preprint,amsmath,amssymb,epsfig]{revtex}

\begin{document}
\title{Scaling of ac susceptibility and nonlinear response in high-temperature superconductors}
 \draft
\author{K.X. Chen, Z.H. Ning, H.Y. Xu, Z. Qi, G. Lu,F.R. Wang and D.L. Yin}
\address{Department of Physics, Peking University, Beijing 100871, China}
\date{\today}
\maketitle

\begin{abstract}
The magnetic ac susceptibility  of high-temperature
superconductors is shown to obey some scaling relations. We try to
analyse this behavior within the framework of a common nonlinear
response function of mixed state.The derived equations for
critical current and ac susceptibility ($\chi(T) $) agree with the
scaling relations of experimental data.

\end{abstract}

\section{Introduction}
One of the most popular means of investigating vortex dynamics in
the high-temperature superconductors(HTS) is the measurement of
the response of the vortex system to ac fields \cite{R1,R2}.Most
experiments with high-$T_{c}$ superconductors (HTS) deal with thin
flat sample in a perpendicular magnetic field. The application of
a time-dependent field $H(t)=H_{0}+h_{ac}e^{-i\omega t}$ to the
sample surface results in an electric-field gradient in the sample
interior ($h_{ac}$ is the ac-field amplitude and $\omega$ the
angular frequency). This gives rise to a shielding current, which
in turn exerts a Lorentz force on the vortices in the sample. The
measurements of such ac response contain much valuable information
about pinning and creep of vortices and turn out to be useful to
test models for describing the ac losses which have to be
carefully characterized and monitored for many applications.

The controversy over the analysis of the ac response is often
noteworthy because of the complicate interplay of hysteretic and
eddy-current losses\cite{R1,R2}. Critical state models like Bean
model and its modifications are often used to analyze the results
where the hysteretic losses dominate. However some problems with
the amplitude dependence of ac responses still remain unsolved.
Several common features of the amplitude dependence of the
in-phase and out-of-phase susceptibilities have been observed for
different kinds of materials \cite{R1,R2,R3,R4,R5,R6}: a) a
parallel shift of the in-phase susceptibility $\chi-T$ curve with
increasing $h_{ac}$ toward lower temperatures is observed; b) the
onset of diamagnetism and dissipation does not appear to depend on
$h_{ac}$ values; c) the out-of-phase peak shifts to lower
temperatures with increasing $h_{ac}$ and broadens in the low
temperature side; d) the absorption peak increases slightly when
$h_{ac}$ increases, In Fig.1 we show the data of high $J_c$ YBCO
bulk material \cite{R5} for an example.

To describe the behavior of the vortex system on a macroscopic
scale, general methods are the Maxwell equations combined with the
materials equation of superconductors $J(E,B,T)$.In present work
we try to analyze the low-frequency ac susceptibility within the
framework of  a common nonlinear response function generally valid
for all type-II superconductors.

In next section we show some widely observed scaling relations of
ac susceptibility experimental data.In section III we introduce
the common nonlinear response function of mixed state. The
equations of critical current and the ac susceptibility are
derived in section IV.A discussion of the connection between the
observed scaling relations and the nonlinear response function is
given in section V.

\section{Scaling behavior of susceptibility}
It is interesting to note that the different measured
$\chi^{''}(T)$ curves in Fig.1 can be represented by a single
curve in Fig.2 when $\chi^{''}(T) - \chi^{''}(T^*)$ and $T-T_{p}$
are scaled by $\chi''_{peak}-\chi^{''}(T^*)$ and $T^{*} - T_{p}$
respectively, with $T^{*}$ the irreversibility point and
$\chi^{''}(T_p)\equiv \chi''_{peak}$ the peak value of the
out-of-phase susceptibility as
\begin{equation}\label{1}
  \frac{\chi^{''}(T)-\chi^{''}(T^*)}{\chi^{''}_{peak}-\chi^{''}(T^*)}=f_{1}(\frac{T-T_p}{T^*-T_p})
\end{equation}

A rather more amazing fact is that the experimental data of
susceptibility from different references \cite{R2, R4, R5} at
different frequencies up to $26 MHz$ can also be superimposed with
the empirical scaling relation
\begin{equation}\label{2}
  \frac{\chi^{''}(T)-\theta(T-T_p)\chi^{''}(T^*)}{\chi^{''}_{peak}-\theta(T-T_p)\chi^{''}(T^*)}
  =f_{2}(\frac{T-T_{p}}{T^{*}-T_{p}})
\end{equation}

 as shown in Fig.3.

The observed $\chi^{''}$ peak position $T_p$ can also be
approximately described by an empirical relation
\begin{equation}\label{3}
  [T_{P}(h_{ac})-T^{*}]^{\alpha}\propto h_{ac}
\end{equation}

as illustrated by the inset of Fig.1,where $T^{*}$ is the limit of
$T_{p}$ as $h_{ac} \rightarrow 0$. This power law shift of $T_p$
to lower values by increasing amplitude $h_{ac}$ can be found for
various kinds of materials. In Fig.4, we summarized some
experimental $h_{ac} \sim T_{p}$ data from different references in
literature.

\section{Nonlinear Response}
The equations that describe the behavior of superconductor on a
macroscopic scale are the Maxwell equations combined with the
materials equation of superconductor $J(E,B,T) $. Various models
in literature suggested different specific forms of the materials
equation. In the case of ideal type-II superconductors with
negligible flux pinning, the material can be characterized by the
linear equation
\begin{equation}\label{4}
  E=\rho_{f}(B,T)J
\end{equation}

with $\rho_{f}\approx\rho_{n}B/B_{c2} $, the flux-flow resistivity
as estimated by Bardeen and Stephen \cite{R7}. On the other hand,
in nonideal type II superconductors with considerable pinning, the
material is described by a set of equations $E=B\times v$,
$v=v_{0}\exp[-U(J)/kT] $or
\begin{equation}\label{5}
  E(J)=J\rho_{f}e^{-U(J, B, T)/kT}
\end{equation}

The activation barrier U depends on $J$ as well as additionally
depends on the temperature T and magnetic field B. Different types
of $U(J)$ have been suggested to approximate the real barrier, for
instance, the Anderson-Kim model \cite{R8} with
$U(J)=U_{c}(1-J/J_{co})$, the logarithmic barrier
$U(J)=U_{c}\ln(J_{co} /J)$ \cite{R9} and the inverse power-law
with $U(J)=U_{c}[(J_{co}/J)^{m}-1]$. \cite{R10,R11,R12}

We find, if one makes a common modification to the different model
barriers $U(J)$ as

\begin{equation}\label{6}
  U(J)\rightarrow U(J_{p}\equiv J-E/\rho_{f})
\end{equation}

then the corresponding modified materials equation
\begin{equation}\label{7}
  E(J)=J\rho_{f}e^{-U(J_{p})/kT}
\end{equation}

leads to a common normalized form as
\begin{equation}\label{8}
  y=x\exp[-\gamma(1+y-x)^{p}]
\end{equation}

with x and y the normalized current density and electric field
respectively. $\gamma $is a parameter characterizing the symmetry
breaking of the pinned vortices system and p is an exponent.

To show the connection of the nonlinear response function Eq. (8)
with the critical- state model $U(J)$, we start from the
expression widely used for flux creep with the logarithmic barrier
\cite{R9},
\begin{equation}\label{9}
  E(J)=\rho_{f}J\exp[-\frac{U_{c}}{kT}\ln(\frac{J_{c0}}{J})]
\end{equation}

Substituting $J_{p}\equiv J-E(J)/\rho_{f}$ for the current density
$J$ in the bracket on the right-hand side of Eq. (9), we get
\begin{equation}\label{10}
  E(J)=\rho_{f}J\exp[-\frac{U_{c}}{kT}\ln\frac{J_{c0}}{J_{p}}]
\end{equation}

The definition of barrier implies $J_{co}\geq J_{p}$. Using the
approximation
\begin{equation}\label{11}
  \ln\eta=\sum_{n=1}^{\infty}\frac{1}{n}(1-\eta^{-1})^{n}\approx
  a(1-\eta^{-1})^{p},(\eta>\frac{1}{2})
\end{equation}

finally we find Eq. (10) in the form
\begin{equation}\label{12}
  \ln(\frac{x}{y})=\gamma(1+y-x)^{p}
\end{equation}

which is the general normalized form of the materials equation Eq.
(8). Here we have
\begin{equation}\label{13}
  \gamma\equiv a\frac{U_{c}}{kT},\quad
  x\equiv\frac{J}{J_{c0}} , \quad
  y\equiv\frac{E(J)}{\rho_{f}J_{c0}}
\end{equation}

In earlier works, this materials equation for type-II
superconductors has also been shown in connection with the
Anderson-Kim model and the inverse power-law $U(J)$
\cite{R13,R14}.

The numerical factor $a$ in the approximation Eq.(11) should be
evaluated with considering the limitation of sample size to the
realistic barrier $U(J)$ as discussed in Refs. \cite{R12,R13,R14}.
Considering this limitation as a cut-off of the series in Eq.
(11), we have $a=\sum_{n=1}^{N_{c}}\frac{1}{n}=C+\ln(N_{c})$

where C is the Euler constant and Nc corresponds to the realistic
cut-off of the series in Eq. (11). Usually a is of the order 2-4.
Ignoring this limitation, one gets from Eq. (10) an even simpler
expression
\begin{equation*}\label{141}
  E(J)=\rho_{f}J(\frac{J_{p}}{J_{c0}})^{U_{c}/kT}
\end{equation*}
or
\begin{equation}\label{142}
  y/x=(x-y)^{\sigma}
\end{equation}

 with $\sigma=U_{c}/kT$, though the latter
can not be used to interpret the case with small barrier and
thermally assisted flux-flow (TAFF). In Fig.5, we show the
numerical solutions of Eq. (8) and Eq. (14) for comparison.

Therefore, the activation barrier $U(J,B,T)$ in Eq.(5) can be
explicitly expressed as
\begin{equation}\label{15}
  U(J,B,T) = U_{c} (B,T) F [J/J_{c0}(B,T)]
\end{equation}

Incorporating it into the commonly observed scaling behavior of
magnetic hysteresis M(H) in superconductors, it can be shown that
$U_{c} (B,T)$ and $J_{co}(B,T)$ in Eq.(15) must take the following
forms \cite{R15}
\begin{eqnarray}\label{16}
  U_{c}(B,T)=\Psi(T)B^{n}\nonumber\\
  J_{c0}(B,T)=\lambda(T)B^{m}
\end{eqnarray}

\section{Critical Current and susceptibility equations}
The nonlinear response function Eq.(8)gives current-voltage
characteristic of the form
\begin{equation}\label{17}
 E(J)=v_{0}Bexp[-\frac{U_{c}(B,T)}{kT}(1+\frac{E(J)}{\rho_{f}J_{c0}(B,T)}-\frac{J}{J_{c0}(B,T)})^{\hat{p}}]
\end{equation}
Where $v_{0}$ is a prefactor with dimension of velocity and
$v_{0}B\approx\rho_{f}J $ as discussed in [13].

Defining the critical current density $J_{c}$ by a certain
criterion of electric field $E_{c}$ as $E(J_{c})\equiv E_{c}$ one
finds from it the expression of the critical surface
\begin{equation}\label{18}
   J_{c}(B,T)=J_{c0}(B,T)[(1-\frac{kT}{U_{c}(B,T)}\ln(\frac{v_{0}B}{E_{c}}))^{\frac{1}{p}}+\frac{E_{c}}{\rho_{f}J_{c0}(B,T)}]
\end{equation}

commonly used for the engineering calculation in applied
superconductivity.In the ac susceptibility measurements we have
$E_{c}=\omega h_{ac}$ and the irreversibility temperature
$T^{*}(B)$ is defined by the condition
\begin{equation}\label{19}
   U_{c}[B,T^{*}(B)]=kT^{*}(B)\ln(\frac{v_{0}B}{E_{c}})
\end{equation}
With this condition,the critical surface equation(18)turns into
the ohmic relation of flux-flow regime as
\begin{equation}\label{20}
  J_{c}[B,T^{*}(B)]=E_{c}/\rho_{f}
\end{equation}
For $T\leq T^{*}(B)$,critical current density can be expressed as
\begin{equation}\label{21}
  J_{c}(B,T)=J_{c0}(B,T)[1-(\frac{T}{T^{*}}\frac{U_{c}(B,T^{*})}{U_{c}(B,T)})^{\frac{1}{p}}+\frac{E_{c}}{\rho_{f}J_{c0}(B,T)}]
\end{equation}

In the case where sample size is much smaller than the wave length
$l$ and the ac amplitude $h_{ac}\ll H_{0}$ one can neglect the
variation of local current density within a period and define two
parameters
\begin{equation}\label{22}
  L_{P}\equiv h_{ac}/J_{c},\quad r\equiv L_{p}/a
\end{equation}
with $a$ the radius of sample,

Substituting Eq.(16)and Eq.(21)to Eq.(22),and considering the
amplitude of electric field induced by $h_{ac}$ $E_{c}=\omega
h_{ac}$,we find the field and temperature dependency
\begin{equation}\label{23}
  r=\frac{h_{ac}}{aJ_{c}}=\frac{h_{ac}}{a}J_{c0}^{-1}(B,T)[1-(\frac{T}{T^{*}}\frac{U_{c}(B,T^{*})}{U_{c}(B,T)})^{1/p}+\frac{c\omega h_{ac}}{\rho_{f}J_{c0}}]^{-1}
\end{equation}
It has been shown by Clem[3],the in-phase and out-of-phase
permeabilities of a type-II superconducting cylinder can be
expressed as
\begin{equation}\label{24}
  \mu'=\mu_{0}'g_{1}(r),\quad \mu''=\mu_{0}'g_{2}(r)
\end{equation}
With the scale function
\begin{eqnarray}\label{25}
   g_{1}(r) & = & r(1-\frac{5}{16}r),0\leq r<1 \nonumber\\
            & = &
            1+\frac{2}{\Pi}[(-\frac{1}{2}+\frac{r}{2}-\frac{5r^{2}}{32})\theta+(-\frac{2}{3r}+1-\frac{7r}{8}+\frac{13r^{2}}{48})]\sin(\theta)+(-\frac{1}{4}+\frac{r}{4}-\frac{r^{2}}{12})\sin2\theta+(-\frac{r}{24}+\frac{r^{2}}{48})\sin3\theta\nonumber\\
            &   &+(-\frac{r^{2}}{384})\sin4\theta,r\geq 1
\end{eqnarray}
where $\theta(r)\equiv\sin^{-1}(r^{-\frac{1}{2}})$ and
\begin{eqnarray}\label{26}
   g_{2}(r) & = & \frac{4}{3\Pi}r(1-\frac{r}{2}),0 \leq r<1\nonumber \\
            & = & \frac{4}{3\Pi}\frac{1}{r}(1-\frac{1}{2r}),r\geq 1
\end{eqnarray}
With a maximum$g_{2}^{MAX}=g_{2}(r=1)=0.21$.

$\mu_{0}^{'}$is the dimensionless differential permeability ,which
increases gradually with decreasing temperature
\begin{equation}\label{27}
  \mu_{0}^{'}\equiv[\frac{dB_{eq}(H)}{dH}]_{H=H_{0}}=\mu_{0}^{'}(H_{0},T)
\end{equation}
  For samples with different geometry we have also the expressions
similar to Eq.(25)and Eq.(26)for susceptibilities $\chi^{'}$and
$\chi''$but with somewhat different specific forms of $g_{1}(r)$
and $g_{2}(r)$than Eqs.(25) and (26)[2].
\section{discussion}
The scaling behavior of ac susceptibility mentioned in section II
can be understood in connection with the nonlinear response
function in section III.Denoting the maximum
$g_{2}^{MAX}(r)=g_{2}(r=r_{p})$,then from Eqs.(22)-(26)we get the
equation for the out-of-phase susceptibility peak position
$T_{p}(B)$in the form
\begin{eqnarray}\label{28}
  r_{p}&=&\frac{h_{ac}}{aJ_{c}(B,T_{p}(B))}\nonumber\\
       &=&\frac{h_{ac}}{a}J^{-1}_{c}(B,T_{p}(B))[1-(\frac{T_{p}U_{c}(B,T^{*}(B))}{T^{*}(B)U_{c}(B,T_{p}(B))})^{1/p}\nonumber\\
       & &+\frac{\omega h_{ac}}{\rho_{f}J_{c0}(B,T_{p}(B))}]^{-1}
\end{eqnarray}
where $U_{c}$and $J_{c0}$can be expressed as [15]
\begin{eqnarray}\label{29}
  U_{c}(B,T)=\Psi(T)B^{n}\propto[T^{*}(B)-T]^{\beta}B^{n}\nonumber\\
  J_{c0}(B,T)=\lambda(T)B^{m}\propto[T^{*}(B)-T]^{\alpha}B^{m}
\end{eqnarray}

Starting from equations (28) and (29) the widely observed scaling
relations Eqs.(1),(2),and (3) can be naturally derived.

In the case of low frequency as in Fig.1,the amplitude of electric
field $E_{c}$ induced by the ac magnetic field $h_{ac}$ is
negligibly small .Thus from Eqs.(23) and (24) one finds
\begin{equation}\label{30}
 \frac{\chi^{''}(T)-\chi^{''}(T^*)}{\chi^{''}_{peak}-\chi^{''}(T^*)}\approx
 \frac{\chi^{''}}{\chi^{peak}}=\frac{g_{2}[r(T)]}{g_{2}^{MAX}}
\end{equation}
using equations (28) and (29) we get the form
\begin{eqnarray}\label{31}
  \frac{\chi^{''}(T)-\chi^{''}(T^*)}{\chi^{''}_{peak}-\chi^{''}(T^*)}\approx[g_{2}^{MAX}]^{-1}g_{2}\{r=r_{p}\frac{J_{c}(B,T_{p}(B))}{J_{c}(B,T)}\}
   =[g_{2}^{MAX}]^{-1}g_{2}\{r=r_{p}[\frac{T^{*}(B)-T_{p}}{T^{*}(B)-T}]^{\alpha}\}\nonumber\\
   =[g_{2}^{MAX}]^{-1}g_{2}\{r=r_{p}[1-\frac{T-T_{p}}{T^{*}-T_{p}}]^{-\alpha}\}
\end{eqnarray}
which is just the scaling relation Eq.(1)

In the case of radio frequency the ac losses due to flux flow is
significant at high temperatures near the irreversibility line.So
the last terms in the right hand sides of equations (21),(23)and
(28) can no longer be omitted.However,the terms with the Heaviside
function $\theta(T-T_{p})$ in the empirical scaling relation
Eq.(2) properly substract these frequency dependent contributions
from the overall critical currents and susceptibilities.Thus,again
we see
\begin{equation}\label{32}
  \frac{\chi^{''}(T)-\theta(T-T_{p})\chi^{''}(T^*)}{\chi^{''}_{peak}-\theta(T-T_{p})\chi^{''}(T^*)}\approx[g_{2}^{MAX}]^{-1}g_{2}\{r=r_{p}[1-\frac{T-T_{p}}{T^{*}-T_{p}}]^{-\alpha}\}
\end{equation}
as the experimental data from different references at different
frequencies up to 26MHz are superimposed in Fig.3.

The amplitude effect relation Eq.(3)can also be well
understood.Since the contribution to critical current from pinning
is dominating at temperature $T=T_{p}$.Omitting the last term in
Eq.(23), one derives from Eqs.(28)and (29)naturally the empirical
relation for the peak position of $\chi^{''}$
\begin{equation}\label{33}
  [T_{p}(h_{ac})-T^{*}]^{\alpha}\propto h_{ac}
\end{equation}

\section{summary}
We find some empirical scaling relations for the ac susceptibility
of high\_temperature superconductors.Based on the analysis of the
nonlinear response function of mixed state we derive the critical
current and susceptibility equations which lead naturally to the
observed scaling behavior.

\acknowledgements{}

This work is supported by the Ministry of Science \& Technology of
China (NKBRSG-G 1999064602) and the Chinese NSF.


\begin{figure*}
\caption{\label{fig1} $\chi^{'}$ and $\chi^{''}$ as functions of
temperature for a YBCO sample with high $J_{c}$ at four values of
$h_{ac}$ (A:11.2 Oe;C:2.2 Oe;D:1.1 Oe) with $h//c$
axis.Inset:relation between $h_{ac}$ and the temperature at the
peak[5].$(f=337Hz)$ }
\end{figure*}

\begin{figure*}
\caption{\label{fig2}Scaling form of the $\chi^{''}(T)$ curves in
Fig.1 with different $h_{ac}$ noted by A,B,C,and D respectively. }
\end{figure*}

\begin{figure*}
\caption{The $\chi^{''}(T)$ curves of high $J_{c}$ YBCO bulk
material,$Yba_{2}Cu_{3}O_{7}$ single crystal and high quality
$Yba_{2}Cu_{3}O_{7}$ films at different frequencies $(337Hz\sim
26Hz )$,can be superimposed when the susceptibility is scaled as
$\frac{[\chi^{''}(T)-\chi^{''}(T^{*})\theta(T-T_{p})]}{[\chi^{''}(T_{p})-\chi^{''}(T^{*})\theta(T-T_{p})]}$
and the temperature is scaled as
$[\frac{(T-T_{p})}{(T^{*}-T_{p})}]$.}A denote the $c^{''}(T)$
curves in Ref.[5] with $f=337Hz$;B denote the $c^{''}(T)$ curves
in Fig.5(b),6(b),8(b) of Ref.[4] with $f=26,0.1,9MHz$;C denote the
$c^{''}(T)$ curves in Fig.2(a),2(b),6(a),6(b) of Ref.[2].
\end{figure*}

\begin{figure*}
\caption{Relation between $h_{ac}$ and $(1-T_{p}/T^{*})$ in the
different experiments.$T_{p}$ is the temperature at the peak of
$\chi^{''}$.$\Box$:YBCO bulk sample [5];$\bullet$:Single crystal
of $Pr_{1.85}Ce_{0.15}CuO_{4-y}$ at $f=111Hz$ and $\mu_{0}H=1T$
[6];$\blacksquare$:Single crystal of $Pr_{1.85}Ce_{0.15}CuO_{4-y}$
at $f=111Hz$ and $\mu_{0}H=0.1T$ [6];$\blacktriangledown$:a disk
(diameter $1mm$) of YBCO film [2];$\blacklozenge$:a rectangle
$(2\times 3mm^{2})$ of YBCO film [2];$\bigcirc$:a ring (width
$50\mu m$) of film YBCO film [2];$\blacktriangle$:a ring (width
$25\mu m$) of YBCO film [2]. }
\end{figure*}

\begin{figure*}
\caption{Numerical solutions of equation (8) (open symbols) and
equation (14) (lines) for comparison.}
\end{figure*}

\end{document}